\documentclass[letter]{aa} 

\usepackage{epsfig}	
\usepackage{graphicx,color}	
\usepackage{amssymb}		
\usepackage{url}		
\usepackage{amsmath}		
\usepackage{rotating}			
\usepackage{float}			
\usepackage{textcomp}		
\usepackage{epstopdf}
\usepackage{dcolumn}
\usepackage{times}
\usepackage{tabularx}
\usepackage{hyperref}
\hypersetup{
    colorlinks,
    citecolor=blue,
    filecolor=blue,
    linkcolor=blue,
    urlcolor=blue,
    menucolor=black}
\usepackage{soul} 
\usepackage[english]{babel}
\usepackage{booktabs}
\usepackage{ulem}

\begin{document}


\title{Flare induced decay-less transverse oscillations in solar coronal loops}
   \author{Sudip Mandal
          \inst{1}
          \and
          Hui Tian\inst{2,3}
          \and
          Hardi Peter\inst{1}
          }

   \institute{Max Planck Institute for Solar System Research, Justus-von-Liebig-Weg 3, 37077, G{\"o}ttingen, Germany \\
              \email{smandal.solar@gmail.com}
         \and
             School of Earth and Space Sciences, Peking University, Beijing, 100871, China
          \and
             Key Laboratory of Solar Activity, National Astronomical Observatories, Chinese Academy of Sciences, Beijing 100012, China
             }

\abstract{

Evidence of flare induced, large-amplitude, decay-less transverse oscillations is presented. A system of multi-thermal coronal loops as observed with the Atmospheric Imaging Assembly (AIA), exhibit decay-less transverse oscillations after a flare erupts nearby one of the loop footpoints. Measured oscillation periods lie between 4.2 min and 6.9 min wherein the displacement amplitudes range from 0.17 Mm to 1.16 Mm. A motion-magnification technique is employed to detect the pre-flare decay-less oscillations. These oscillations have similar periods (between 3.7 min and 5.0 min) like the previous ones but their amplitudes (0.04 Mm to 0.12 Mm) are found to be significantly smaller. No phase difference is found among oscillating threads of a loop when observed through a particular AIA channel or when their multi-channel signatures are compared. These features suggest that the occurrence of a flare in this case neither changed the nature of these oscillations (decaying vs decay-less) nor the oscillation periods. The only effect the flare has is to increase the oscillation amplitudes. 
}

   \titlerunning{Flare induced decay-less transverse oscillations in solar coronal loops}
   \authorrunning{Sudip Mandal et al.}
   \maketitle

\newpage
\section{Introduction}
The solar corona, a highly structured million Kelvin hot and dynamic upper atmosphere of the Sun, accommodates a wide variety of magnetic structures including coronal loops. These loops act as wave guides for a range of magnetohydrodynamical (MHD) waves that are observed in the corona \citep{2005LRSP....2....3N}. One such wave mode is the transverse or kink oscillation. Since the time of its discovery \citep{1999ApJ...520..880A,1999Sci...285..862N}, kink waves have been routinely identified as transverse loop displacements in extreme ultraviolet (EUV) images (see \citet{2009SSRv..149..199R,2020ARA&A..58..441N} for detailed reviews). Over various observations, it has been found that these kink waves are often triggered by a nearby transient event such as a flare \citep{2002SoPh..206...69S}, or a jet \citep{2015A&A...577A...4Z}, or by a coronal mass ejection (CME) \citep{2012ApJ...751L..27W}. Furthermore, once triggered, these waves often die down rapidly (within couple of wave periods). This rapid decay of the wave amplitude has now been established as a key observational feature of kink waves \citep{2002SoPh..206...99A}. In MHD wave theory, these oscillations are described as the fast magneto-acoustic wave mode which is weakly compressible in the long wavelength limit \citep{1983SoPh...88..179E,1984ApJ...279..857R,2008ApJ...676L..73V}. Further, the observed rapid damping is explained through a mechanism known as the resonant absorption \citep{2002A&A...394L..39G,2002ApJ...576L.153O,2011SSRv..158..289G}.

Over the last decade or so, the availability of high-resolution and high-cadence EUV data, such as from the Atmospheric Imaging Assembly \cite [AIA;][]{2012SoPh..275...17L}, has significantly improved our understanding of kink oscillations. In-fact, it has also led to the discovery of a new kind of transverse oscillations, termed as `decay-less' kink oscillations \citep{2012ApJ...759..144T,2012ApJ...751L..27W,2013A&A...552A..57N}. These are small amplitude ($\leq$0.5Mm) transverse oscillations with periods between 3-7 min and show no apparent decay in their amplitudes over a large number of wave periods. These waves are rather omnipresent in active region coronal loops and seem to have no apparent connection with any transient events such as a flare\footnote{Although almost all decayless oscillations have no connection to eruptive events, \citet{2012ApJ...751L..27W} reported an event triggered by a CME.} \citep{2015A&A...583A.136A}. This is particularly different from the traditionally observed rapidly decaying kink waves which are always associated with an impulsive driver. Recent theoretical studies as well as MHD simulations on decay-less kink waves, have hypothesized that these oscillations are driven either by the random footpoint motions \citep{2013A&A...552A..57N,2016A&A...591L...5N,2020A&A...633L...8A} or through a monoperiodic driver \citep{2019ApJ...876..100A}. However, an exact match with the observed wave properties is yet to be achieved.

In this letter, we present a unique event where several multi-thermal coronal loops exhibit decay-less transverse oscillations that are triggered by a nearby flare. In Section 2 we describe the event and the data, whereas the Section 3 outlines the results. Finally, we conclude by summarizing our findings in Section 4.


\begin{figure*}[!htb]
\centering
\includegraphics[width=0.95\textwidth,clip,trim=0cm 0cm 0cm 0cm]{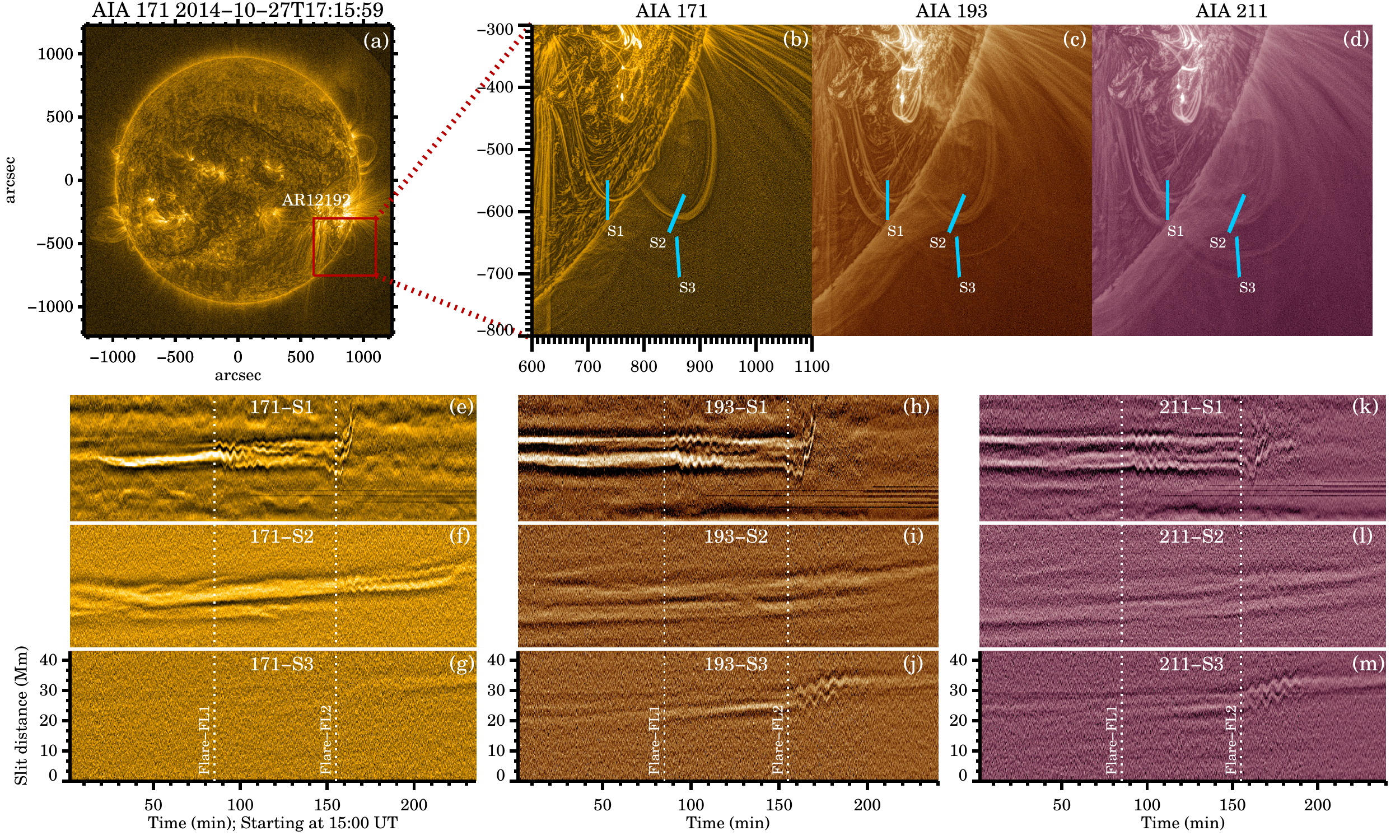}
\caption{ 
Overview of oscillating loops and their temporal evolution. {\it Panel-a:} A representative full-disc image of the Sun highlighting the location of our event by the red rectangular box. Images of this boxed region from 171~{\AA}, 193~{\AA} and 211~{\AA} channels are presented in {\it Panels-b,d}, respectively. Blue straight lines, in each panel, mark the locations of the three artificial slits ($\rm{S1}$, $\rm{S2}$, $\rm{S3}$) which are used to generate the space-time (XT) maps. Generated XT maps are shown in {\it panels-e,m}. Two vertical dotted lines, in every panel, mark the onset time of the two flares.
}

\label{fig1}
\end{figure*}

\section{Data and event details}\label{data_details}

In this study, we use high-resolution extreme ultraviolet (EUV) imaging data from the Atmospheric Imaging Assembly \cite[AIA;][]{2012SoPh..275...17L}, onboard Solar Dynamics Observatory \cite[SDO;][]{2012SoPh..275....3P}. Images from three different EUV passbands, namely the 171~{\AA}, 193~{\AA} and 211~{\AA} channels have been analyzed. The data we present here cover 240 minutes (staring from 15:00 UT on 27-10-2014) with a cadence of 12~s. The pixel scale, in each direction, is $\approx$0.6$^{\prime\prime}$.

The event occurred within a system of coronal loops that are rooted inside the active region AR12192. This active region was located near the South-West limb of the Sun as shown in In Fig.~\ref{fig1}a. As is the case with any typical AIA EUV image, the off-limb coronal loops in our data appear to be fuzzy. This happens primarily due to the diffuse nature of the EUV emission. In order to accurately infer the nature of the observed oscillations, it is therefore important that we correctly identify the individual loops (and their strands). To bring out such finer details, we use the `Multi-Gaussian Normalization' (MGN) image processing technique as developed by \citet{2014SoPh..289.2945M}. As described in that paper, the MGN method basically normalizes an image at multiple spatial scales and also flattens the noisy regions, highlighting the small scale details. All the data from 171~{\AA}, 193~{\AA} and 211~{\AA} channels are processed using this MGN method and the images presented in Panels~\ref{fig1}(b-d) are some of the examples of this processing. A quick look at these panels reveals two important aspects about the structure of these loops. Firstly, each loop is made up of multiple individual strands and secondly, although most of these loops can visibly be traced in all three AIA channels, their contrasts are significantly different across these channels. In fact, some of these loops are even absent in the 171~{\AA} channel and only visible in the hotter 193~{\AA} or 211~{\AA} channel, implying that these structures are probably 1.5~$\rm{MK}$ to 2~$\rm{MK}$ hot and thus, not prominent in the 171~{\AA} channel. All of these characteristics are basically a manifestation of the multi-thermal nature of the loop plasma.


\section{Results and discussion}

From the movie (available online) we see that, around 16:30 UT, a flare erupts close to the left footpoint of this loop system we described in Sect.~\ref{data_details} and subsequently triggers oscillations in the nearby loops. These oscillations continue to exist untill about 17:40 UT, when we see a second flare erupting at the exact same location as the previous one. This second flare is visibly more intense and as a result, it completely disrupts the nearby loops that were oscillating till then. However, some of the far away loops survive this flare outburst and they start oscillating thereafter.

\subsection{AIA light curves}

\begin{figure}[!htb]
\centering
\includegraphics[width=0.45\textwidth,clip,trim=0cm 0cm 0cm 0cm]{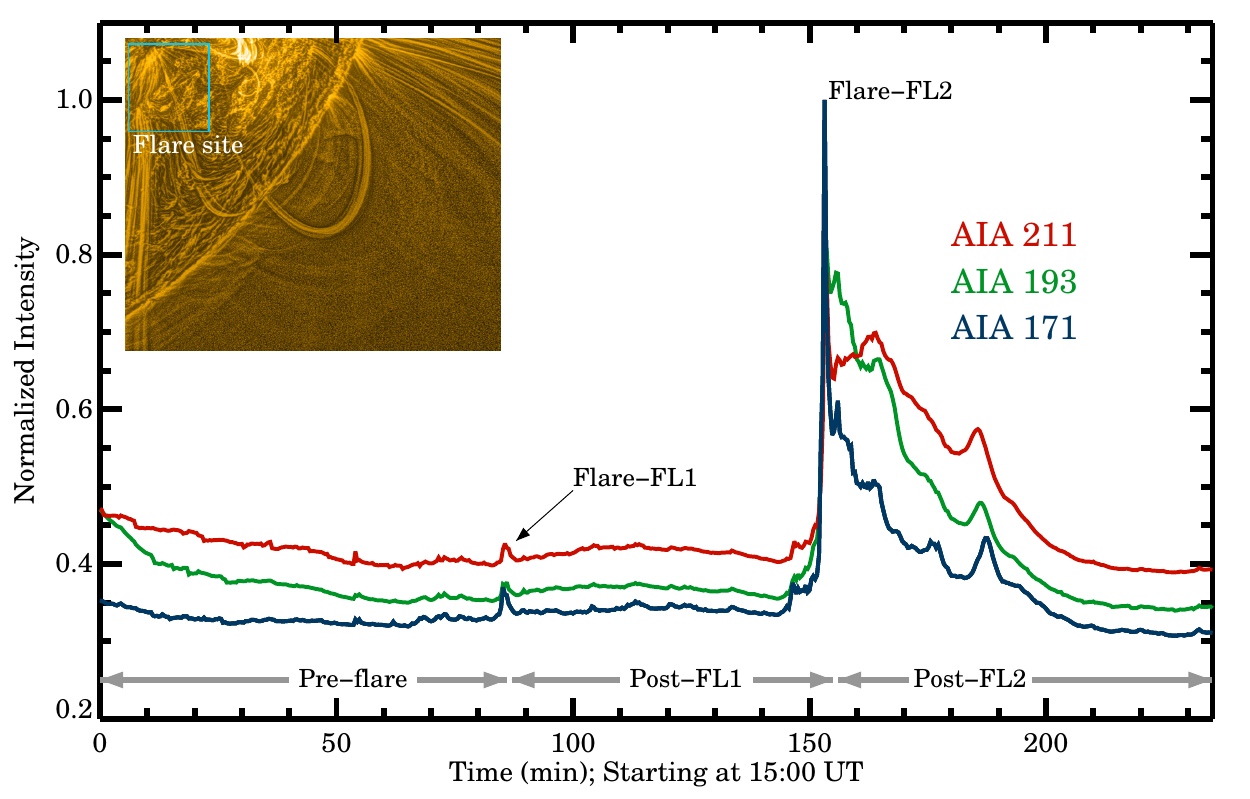}
\caption{
Light curves from three different AIA channels. {\it The inset} shows the flaring site outlined by the blue box drawn ontop of the 171~{\AA} image. Normalized intensity curves extracted from this boxed region are shown in {\it the main panel}. The two observed flares are also highlighted in the panel.
}
\label{fig6}
\end{figure}

We analyze the AIA light curves to better understand the flar-
ing activities that we see near the loop footpoint. The blue box, drawn ontop of the 171~{\AA} image (Fig.~\ref{fig6}a), outlines our region of interest that includes the flaring site as well as the left loop footpoint. Light curves, for each of the three AIA channels, are then generated by averaging the intensity\footnote{These intensities correspond to the original AIA measurements i.e before the MGN processing.} values over this box. These curves are further normalized by their corresponding maximas and the final values are presented in Fig.~\ref{fig6}b. From the plot, we find that there are no significant activities in any of these AIA channels during initial times. The first flare, FL1, erupts around t$\approxeq$85 min and can be seen as near simultaneous intensity enhancements in all three light curves. This flare lasts only about 3 minutes and another quiet period follows thereafter. Things however change rapidly around t$\approxeq$150 min, marking the onset of the second flare, FL2. This flare is significantly stronger than FL1 (the AIA intensities, in all three channels, saturate during the peak of this flare). Additionally, there are multiple small intensity enhancements riding over these light curves during the decay phase of FL2. Overall, using these characteristics, we can divide our data into three activity phases i.e., pre-flare phase, post-FL1 phase and post-FL2 phase.

\begin{figure*}[!htb]
\centering
\includegraphics[width=0.95\textwidth]{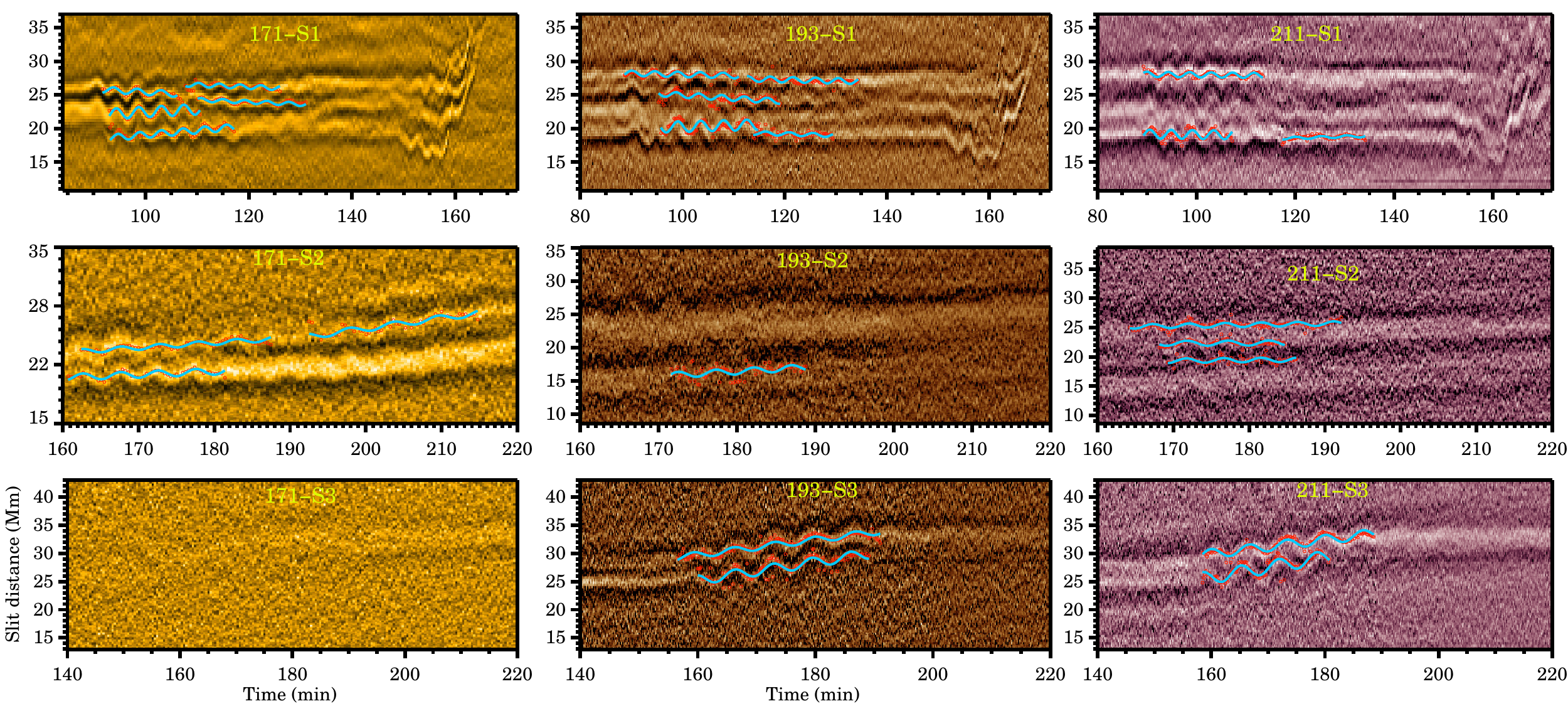}
\caption{
Fits to the individual oscillations. The red points highlight the skeletons of the individual oscillating structures, whereas the best fitting sinusoidal function (Eq.~\ref{equation1}) is shown by the blue curves. See Sect. 3.2.1 for more details.
}
\label{fig2}
\end{figure*}

\subsection{Space-time maps}

To generate the space-time (XT) maps, we use multiple artificial transverse slits. The blue straight lines in Fig.~\ref{fig1}b-d, mark the locations of these slits ($\rm{S1}$, $\rm{S2}$, $\rm{S3}$) wherein the corresponding XT maps are shown in Panels~\ref{fig1}e-m. In order to increase the intensity signal, we choose each of these slits to be 7 pixels wide for $\rm{S1}$ and $\rm{S2}$ and 9 pixels wide for $\rm{S3}$. The final XT maps are then generated by averaging over their corresponding slit widths. Furthermore, for a chosen loop, we generate multiple XT maps by placing a series of slits and then manually select the best one where the oscillation signatures appear to be clearest by eye estimation. 

\subsubsection{Post-flare oscillations}

We first focus our attention to the oscillations that appear in the XT maps post FL1 and FL2. In all of these maps (Fig.~\ref{fig1}e-m), one can immediately notice the decay-less nature of most of these oscillations\footnote{We do note that the oscillation in the central thread in Fig.~\ref{fig1}e appears more like a slowly decaying one.} i.e no apparent decay in their amplitudes even after several wave periods. In extreme cases (e.g., Fig.~\ref{fig1}e,f,h,k) the amplitudes, in fact, do not show any decaying signature even after 50 minutes post FL1. We now analyze each of these individual XT maps. In case of the maps for slit $\rm{S1}$, oscillations appear immediately after FL1 (t$\approxeq$88 min) and are prominently visible in all three AIA channels. On the other hand, oscillations in $\rm{S2}$ only start after FL2 (t$\approxeq$153 min). This time, the oscillatory signatures are best seen in 171~{\AA} channel and are faintly visible in other two channels. The scenario however reverses in $\rm{S3}$ maps where we find clear oscillations (which again appear only after FL2) in the hotter 193~{\AA} and 211~{\AA} channels and a significantly weaker signal in 171~{\AA} channel. 

Next, we proceed onto fitting these individual oscillations. For each oscillating strand, we first determine its `skeleton' by fitting a Gaussian along the transverse direction of these maps. The centers of these fitted Gaussians are then grouped together to constitute a skeleton. In each panel of Fig.~\ref{fig2}, we highlight these centers by the red circles. A combination of a sine function and a linear trend 
\begin{equation}
\label{equation1}
\centering
 y(t)=A* \left (\sin{\dfrac{2\pi t}{\rm{P}}+\phi} \right)+c_1 t+c_0
\end{equation}

 is then fitted to these individual skeletons (belonging to a particular thread). Here $\rm{A}$ is the oscillation amplitude, $\rm{P}$ represents the period, $\phi$ is the phase and  c$_{0}$ and c$_{1}$ are constants.

The final fits to the detected oscillations are overplotted by the blue curves in different panels of Fig.~\ref{fig2}. The fitted periods ($\rm{P}$) and amplitudes ($\rm{A}$) of all detected oscillations are listed in Table~\ref{table1} (Post-FL1 and Post-FL2 colums). From the table, we note that the wave periods lie between 4.2 min and 6.9 min, with the majority of them being close to 4.5 min. This is consistent with the previous reports of decay-less oscillations \citep{2013A&A...560A.107A,2015A&A...583A.136A}. The amplitude values, on the other hand, show a broader distribution with values between 0.17 Mm and 1.16 Mm. Interestingly, the majority of these $\rm{A}$ values are considerably larger than typical amplitude values we see in decay-less oscillations \citep{2015A&A...583A.136A}.

\subsubsection{Pre-flare oscillations}

Previous reports of decay-less kink waves highlighted the fact that these oscillations are omnipresent in coronal loops and have no apparent association with flares \citep{2012ApJ...759..144T,2015A&A...583A.136A}. Thus, it is natural for us to look for the signatures of such oscillations in the pre-flare phase data of our event. As shown earlier, the data before the first flare (i.e before 16:29 UT) is considered as the `pre-flare' data in this case.

\begin{figure}[!htb]
\centering
\includegraphics[width=0.50\textwidth,clip,trim=0cm 0cm 0cm 0cm]{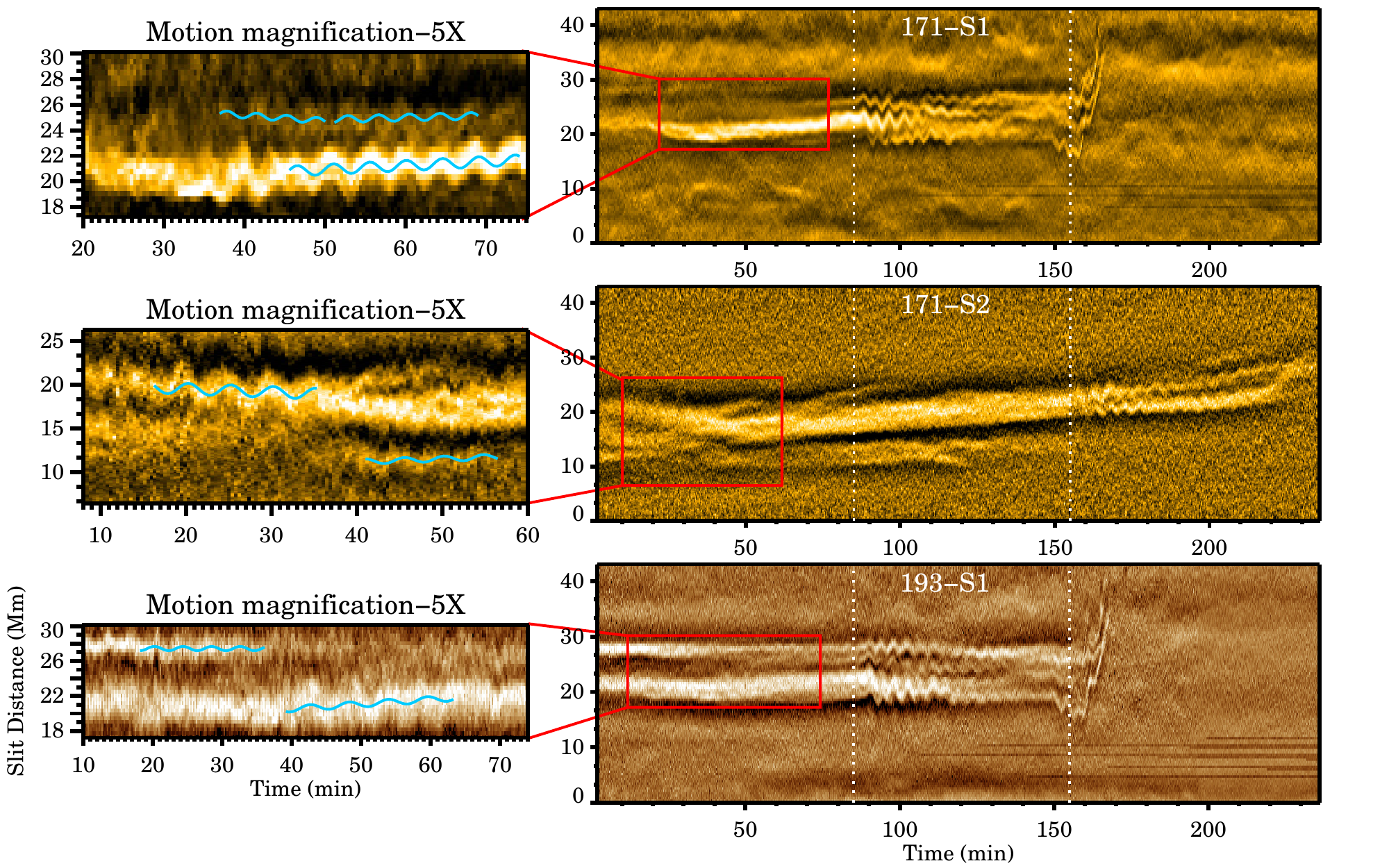}
\caption{Pre-flare oscillation maps. {\it Right panels (top to bottom):}  Space-time (XT) maps from slits 171-$\rm{S1}$, 171-$\rm{S2}$ and 193-$\rm{S1}$, respectively. In each case, a red rectangular box outlines the pre-flare phase of the data that is used to generate the corresponding motion magnified map. {\it Left panels} shows the motion magnified maps for the three slits mentioned before. The blue curves show the best fitting sinusoidal function onto the detected oscillatory patterns. See Sect. 3.2.2. 
}
\label{fig3}
\end{figure}

A careful inspection of each of the original XT diagrams (panels~\ref{fig1}e-m) reveals that, out of all these maps, definite signatures of such small amplitude, pre-flare oscillations are only visible in the XT maps of slit $\rm{S1}$ of 171~{\AA} and 193~{\AA} channels and rather faintly, in the XT maps of slits $\rm{S2}$ of 171~{\AA} channel (see Fig.~\ref{fig1}e,g,h). Although these oscillatory patterns can be traced visually, their signals are not sufficient to perform any reliable fit. Thus, we use a `motion magnification' technique \citep{2016SoPh..291.3251A} to enhance these oscillation signals. This technique uses a dual-tree complex wavelet transform to amplify the signals of transverse motions that are present within an image sequence. The right panels of Fig.~\ref{fig3} show the original (non-magnified) XT maps, wherein the pre-flare motion magnified XT maps are shown in the left panels. We tested different magnification factors ($\rm{K}$) between 3 to 10 and after visual inspection, settled on a value of $\rm{K}$=5 to best enhance the oscillation signal.

With these magnified XT maps, we can now better identify these pre-flare oscillations. Although the 193~{\AA} oscillations are somewhat blurry, the 171~{\AA} ones are prominent and their decay-less nature is also very evident. An interesting feature to note here is the shape of these oscillatory signals which appears to be more pointy or triangle-shaped than sinusoidal. Similar characteristics were also seen in the examples presented in \citet{2013A&A...560A.107A,2016ApJ...830L..22A}. We measure the wave periods and amplitudes in the same way as discussed earlier and the best fits to the observed oscillations are shown by the blue curves in Fig.~\ref{fig3}. Derived period and amplitude values are listed in the pre-flare columns of Table~\ref{table1}. Looking at this table, we notice that the pre-flare periods are similar to those of the post-flare values. At the same time, the pre-flare amplitudes are significantly smaller (approximately 3 to 5 times) than the post-flare ones. Interestingly, our pre-flare amplitudes match well with the amplitude values reported in previous studies of small amplitude decay-less oscillations \citep{2015A&A...583A.136A}. More discussion on this is presented in the next section.

\newcommand*{\mystrut}{\rule{0pt}{1.5ex}}
\begin{table}[!h]
\caption{ Fitted parameters of individual wave threads }  
\label{table1}
\resizebox{\columnwidth}{!}{%
    \begin{tabular} {@{}llcccccc@{}}

       \toprule
     &   & \multicolumn{2}{c}{Pre-flare}   &   \multicolumn{2}{c}{Post-FL1}      & \multicolumn{2}{c}{Post-FL2}                                      \\  \cmidrule(lr){3-4} \cmidrule(lr){5-6} \cmidrule(lr){7-8}
     \mystrut   Slit  &  Channel     & \multicolumn{1}{c}{P (min)} & \multicolumn{1}{c}{A (Mm)} & \multicolumn{1}{c}{P (min)} & \multicolumn{1}{c}{A (Mm)} & \multicolumn{1}{c}{P (min)} & \multicolumn{1}{c}{A (Mm)} \\ \midrule


         &     & 3.7 & 0.04 & 4.2 & 0.33 &  --  & --  \\
         &     & 4.5 & 0.08 & 4.5 & 0.28 &  --  & --  \\
     S1  & 171 & 3.8 & 0.06 & 4.6 & 0.80 &  --  & --  \\
         &     & --  &  --  & 4.4 & 0.50 &  --  & --  \\
         &     & --  &  --  & 4.8 & 0.45 &  --  & --  \\
     \hline
         &     & 4.4 & 0.05 & 4.4 & 0.41  &  -- & --  \\
         &     & 5.6 & 0.07 & 4.4 & 0.38  &  -- & --  \\
     S1  & 193 & --  &  --  & 4.5 & 0.80  &  -- & --  \\
         &     & --  &  --  & 4.5 & 0.17  &  -- & --  \\
         &     & --  &  --  & 4.6 & 0.27  &  -- & --  \\
         &     & --  &  --  & 5.8 & 0.36  &  -- & --  \\
    \hline
         &     & --  &  --  & 4.3 & 0.39  &  --  & --   \\
     S1  & 211 & --  &  --  & 4.3 & 0.66  &  --  & --   \\
         &     & --  &  --  & 5.2 & 0.18  &  --  & --   \\
    \hline
         &     & 4.6 & 0.06 & --  &  --   &  5.0 & 0.38   \\
     S2  & 171 & 5.0 & 0.12 & --  &  --   &  5.1 & 0.26   \\
         &     & --  &  --  & --  &  --   &  5.9 & 0.33   \\
    \hline
     S2  & 193 & --  &  --  & --  &  --   &  4.8 & 0.45  \\
    \hline
         &     & --  &  --  & --  & --    & 4.8  & 0.37  \\
     S2  & 211 & --  &  --  & --  & --    & 5.4  & 0.44  \\
         &     & --  &  --  & --  & --    & 4.9  & 0.39  \\
    \hline
     S3  & 171 & --  &  --  & --  &  --   &  -- & --     \\
    \hline
     S3  & 193 & --  &  --  & --  &  --   &  6.9 & 0.55  \\
         &     & --  &  --  & --  &  --   &  6.6 & 0.87  \\
    \hline
     S3  & 211 & --  &  --  & --  & --    & 6.7  & 0.83   \\
     --  & -- & --  &  --  & --  & --    & 6.7  & 1.16   \\
    \hline
\end{tabular}
}
\footnotesize{Note: Typical fitting error in $\rm{P}$ is $\pm$0.1 minute, whereas the same in $\rm{A}$ is $\pm$0.02 $\rm{Mm}$.}
\end{table}

\subsubsection{Comparison between the pre-flare and post-flare oscillations}

\begin{figure}[!htb]
\centering
\includegraphics[width=0.45\textwidth,clip,trim=0cm 0cm 0cm 0cm]{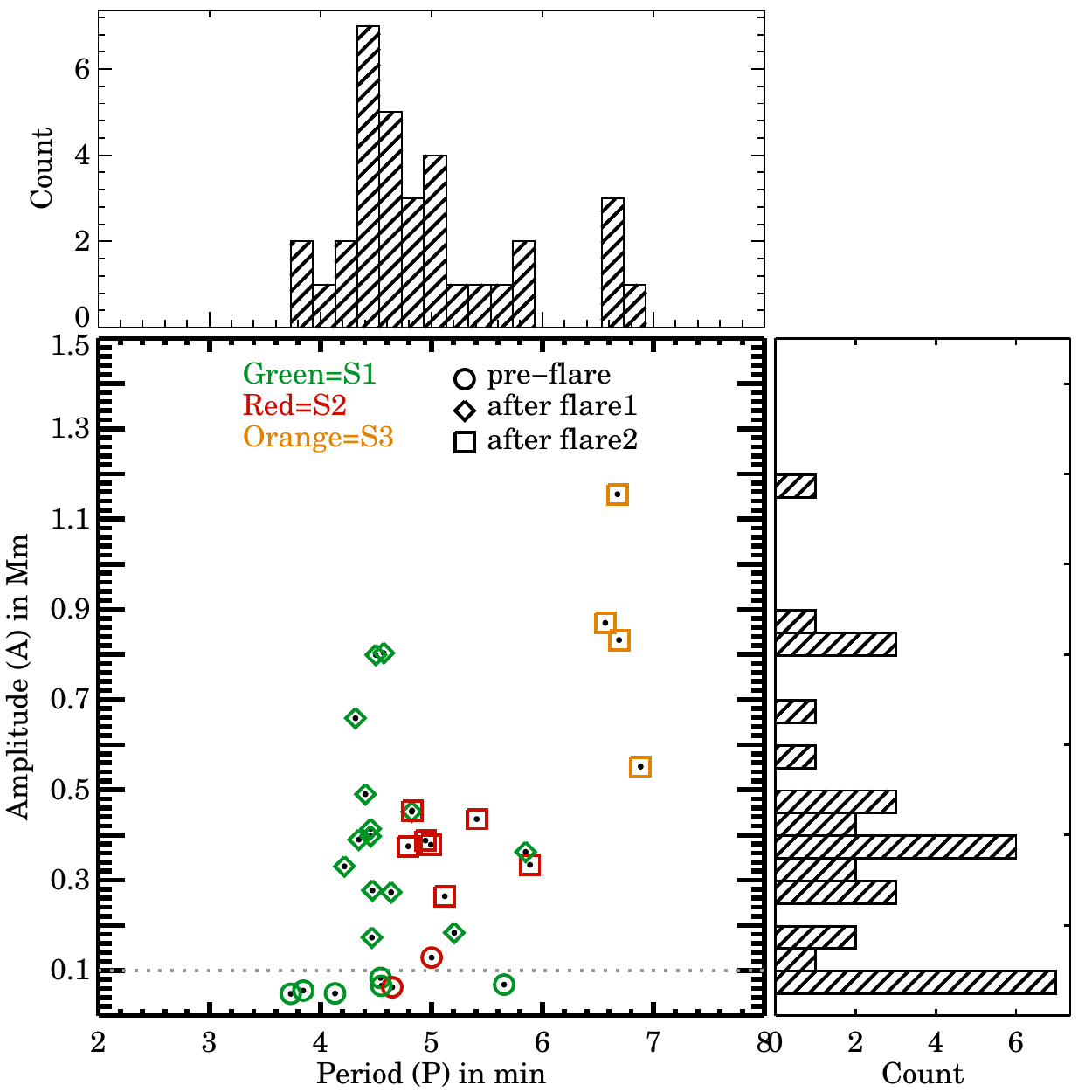}
\caption{
Period-amplitude diagram. Colors (green, red, orange) indicate the slits ($\rm{S1}$, $\rm{S2}$, $\rm{S3}$) from which these oscillations originate, whereas the symbols (circle, diamond, square) illustrate the epoch (pre-flare, post-FL1, post-FL2) of these oscillations. The dotted horizontal line in the plot mark the $\rm{A}$=0.1 Mm boundary.
{\it Top and side panels} show the histograms of period ($\rm{P}$) and amplitude ($\rm{A}$) values.
}
\label{fig4}
\end{figure}

In the discussion of our study, so far we have described that these decay-less oscillations continue to exist in a loop despite multiple flaring events occurring nearby. In order to better understand the impact of these flares on the observed oscillations, we compare here the pre-flare and post-flare wave properties. Fig.~\ref{fig4} shows the period-amplitude diagram that is constructed using all the individual oscillations that are listed in Table~\ref{table1}. In this plot, each data-point is marked with a color that indicates its slit of origin whereas the associated symbol represents its epoch of occurrence. From the period histogram (top panel of Fig.~\ref{fig4}), we notice that the oscillation periods (both, pre and post flare ones) are mostly restricted between 4 to 6 min, although the $\rm{S3}$ periods are grouped near $\rm{P}$=7 min. These longer periods occur only in the space-time plots of $\rm{S3}$. This is clear, because 
 this particular slit, $\rm{S3}$, was placed onto the large distant loop and the kink period ($\rm{P}$) increases linearly with loop lengths \citep{2015A&A...583A.136A}. The distribution of wave amplitudes ($\rm{A}$) show some interesting features. All the pre-flare values are either below or very close to 0.1 Mm, whereas post-FL1 $\rm{A}$ values lie between 0.15 Mm and 0.8 Mm, with majority of them clustering around 0.4 Mm. Thus, on an average, the post-FL1 oscillations have 5 times larger amplitudes than the pre-flare ones, although the wave periods mostly remained unchanged between these two epochs. Looking at this trend, one would be inclined to conclude that this enhancement in wave amplitudes is directly proportional to the strength of the triggering flare. However, when we include the post-FL2 oscillations, such a relation does not manifest well. For example, the post-FL2 $\rm{S2}$ amplitudes are similar to that of the post-FL1 $\rm{S1}$ oscillations whereas, the post-FL2 $\rm{S3}$ amplitudes are bigger. Hence, oscillations of different amplitudes are getting generated by a single flaring event. In the same context, we again recall that the flare FL2, which triggered these oscillations, is a significantly stronger flare than FL1. From this we can conclude that the occurrence of a flare in this case neither changed the nature of these oscillations nor the oscillation periods, but increased their amplitudes.

\subsubsection{Multi-wavelength aspect}

\begin{figure}[!htb]
\centering
\includegraphics[width=0.40\textwidth,clip,trim=0cm 0cm 0cm 0cm]{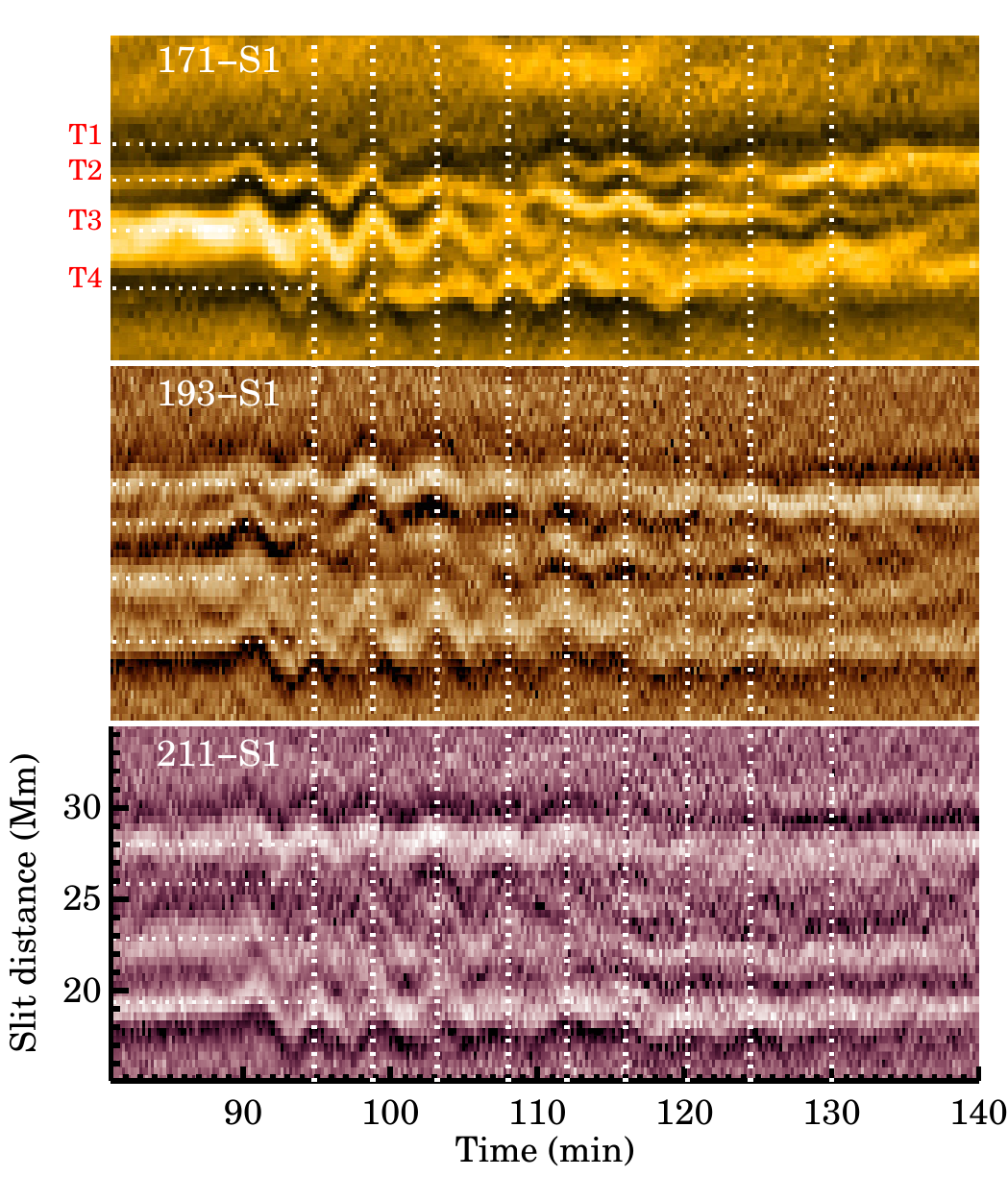}
\caption{ Multiwavelength view of the $\rm{S1}$ oscillations. Vertical lines mark the positions of the wave maximas as seen in 193~{\AA} channel. The horizontal lines highlight the initial locations of the individual loop-threads (T).
}
\label{fig5}
\end{figure}

In this section, we examine the multi-wavelength properties of these oscillations. To do this, we select the time-distance maps from slit $\rm{S1}$ as the oscillation signals in these maps are prominently visible in all three AIA channels. 
Through different panels of Fig.~\ref{fig5}, we show the post-FL1 period of the $\rm{S1}$ maps from 171~{\AA}, 193~{\AA}, and 211~{\AA}, respectively. Further, the individual threads ($\rm{T}$'s) are also marked with dotted horizontal lines. By following these threads, we find that these oscillatory patterns, in all three AIA channels, start exactly at the same time. Additionally, we also notice that, although these oscillations appear more sharper (i.e with better contrasts) in the beginning, they rather become diffuse at later times. This is possibly an artifact of the movements of these individual threads. Moreover, a closer inspection also reveals that the thread $\rm{T1}$, which is visible throughout in 193~{\AA} and 211~{\AA} channels, become visible in 171 channel only after t=110 min. On the other hand, the thread $\rm{T2}$ only appears in channels 171~{\AA} and 193~{\AA} and is absent in 211~{\AA} channel. These are basically the manifestations of the thermal structuring of the loop plasma.
 
Some of the earlier studies \citep[e.g.][]{2012ApJ...751L..27W,2016ApJ...830L..22A} had found a phase difference between such oscillations when observed simultaneously through two different AIA channels. Hence, we also look for such signatures in our data. In Fig.~\ref{fig5}, we mark the the positions of the oscillation maximas at different wave cycles. From the plot, we see that the oscillation maximas appear simultaneously across all AIA channels. Further, this in-phase behavior persists even after multiple wave periods. Hence, we conclude that these oscillating threads, across multiple AIA channels, are moving in phase. Additionally, we found that every thread within a given space-time map, also exhibits synchronous oscillation with all other threads in that map and again, this behavior remains unchanged during the entire lifetime of these oscillations. The in-phase oscillations seen in different AIA channels are consistent with the finding by \citet{2012ApJ...759..144T}, where these authors found in-phase oscillations in different spectral lines observed with Hinode/EIS\footnote{Extreme-ultraviolet Imaging Spectrometer}.

\section{Summary and Conclusion}

In this letter, we study the properties of decay-less transverse oscillations that are triggered by a nearby flaring event. These oscillations appear within a system of coronal loops that are rooted inside an active region. 

\begin{enumerate}

\item
The observed oscillations are transverse in nature and they last for three to ten wave periods without any measurable decay in their amplitudes. These are basically the signatures of decay-less kink waves. The wave periods range between 4.2 min and 6.9 min whereas the wave amplitudes lie within 0.17 Mm and 1.16 Mm.\\

\item
The observed decay-less oscillations are generated by a nearby flare, FL1. Once triggered, these waves continue to exist untill another nearby flare, FL2, disrupts the oscillating loops. Multiwavelength analysis reveals a synchronous appearance of these oscillatory patterns in different AIA channels (171~{\AA}, 193~{\AA}, 211~{\AA}). In any given AIA channel, all loop threads oscillate in phase. Similar in-phase oscillation is also seen across different AIA channels.\\

\item
Small amplitude, pre-flare, decay-less transverse oscillations are also present in our data. These waves exist within the same loop structures in which we previously noted the post-flare oscillations. After classifying each oscillation according to it appearance in the data (i.e pre-flare or post-flare), a pattern emerges. \\

Our main finding is that the post-flare oscillations are found to have 5 to 7 times bigger amplitudes than the pre-flare ones, whereas the oscillation periods, before and after the flare, remained similar. 

\end{enumerate}

The mechanism that drives these decay-less transverse oscillations is still not fully understood. In case of small amplitude pre-flare cases, some authors suggested random footpoint driving (caused by supergranular motions) to be the source \citep[e.g.][]{2013A&A...552A..57N,2016A&A...591L...5N} . The wide range of observed wave periods however, does not match the driver characteristics. In a different proposal, \citet{2016ApJ...830L..22A} argued that the Kelvin-Helmholtz (KH) instabilities near the loop boundaries can potentially mimic the observed decay-less features. Interestingly, the post-flare oscillations that we have found in case of $\rm{S2}$ of 171${\AA}$ channel (see Fig~\ref{fig1}f), look very similar to the synthesized XT maps that were presented in Fig.3 of \citet{2016ApJ...830L..22A}. However, the in-phase oscillations across three AIA channels as seen in our data, would not be consistent with this mechanism. Hence, it is possible that the oscillations that we see in this study are probably governed by a combination of all these individual mechanisms i.e., an impulsive driver (the flares), a perpetual driver (the footpoint motions) and the KH rolls. With the current spatial (and temporal) resolution of modern EUV imaging data, it is rather difficult to identify (and quantify) the individual contributions of each of these mechanisms. Nevertheless,  \citet{2016ApJ...830L..22A} showed how a change in spatial resolution can affect the appearance of these oscillations in EUV images and in this context, future observations from the Extreme Ultraviolet Imager (EUI), on board Solar Orbiter, will be extremely useful in de-entangling certain aspects of these oscillations.

\section{Acknowledgment}
We thank the reviewer for his/her encouraging comments. S.M. acknowledges Prof. Dipankar Banerjee for his help in understanding the data. H.T. is supported by the Strategic Priority Research Program of CAS (grant XDA17040507) and NSFC grant 11825301. The AIA data used here is courtesy of the SDO (NASA) and AIA consortium. The authors would also like to acknowledge the Joint Science Operations Center (JSOC) for providing the AIA data download links.



 \bibliographystyle{aa}
 \bibliography{ref_decayless}
\end{document}